\documentclass[twocolumn,secnumarabic,amssymb, nobibnotes, aps, prl, noeprint]{revtex4-1}

\usepackage{graphicx}
\usepackage{natbib}
\usepackage{textgreek}
\usepackage{siunitx}
\usepackage{graphicx}
\usepackage[utf8]{inputenc}
\usepackage{amsmath}
\usepackage{xcolor}

\newcommand{\HSd}{MoS$_2$/WS$_2$~}
\newcommand{\HS}{MoS$_2$/WS$_2$}
\newcommand{\molyd}{MoS$_2$~}
\newcommand{\moly}{MoS$_2$}
\newcommand{\wolyd}{WS$_2$~}
\newcommand{\woly}{WS$_2$}
\newcommand{\percent}[1]{\SI{#1}{\percent}}

\begin{document}

\title{Hybridized indirect excitons in \HSd heterobilayers
}

\author{Jonas Kiemle$^{1}$, Florian Sigger$^{1,2}$, Michael Lorke$^{3,4}$, Bastian Miller$^{1,2}$, Kenji Watanabe$^5$, Takashi Taniguchi$^5$, Alexander Holleitner$^{1,2}$, Ursula Wurstbauer$^{1,2,6}$}

\affiliation{1 Walter Schottky Institut and Physics Department, Technical University of Munich, Am Coulombwall 4a, 85748 Garching, Germany.       
	\\
	2 Nanosystems Initiative Munich (NIM), Schellingstr. 4, 80799 München, Germany. 
	\\
	3 Bremen Center for Computational Materials Science, University of Bremen, Am Fallturm 1, 28359 Bremen, Germany
	\\
	4 Institute for Theoretical Physics, University of Bremen, 28359 Bremen, Germany
	\\
	5 National Institute for Materials Science, Tsukuba, Ibaraki 305-0044, Japan.
	\\
	6 Institute of Physics, Westfälische Wilhelms-Universität Münster, Wilhelm-Klemm-Str.10, 48149 Münster, Germany.}
\date{\today}
\begin{abstract}
Ensembles of indirect or interlayer excitons (IXs) are intriguing systems to explore classical and quantum phases of interacting bosonic ensembles. IXs are composite bosons that feature enlarged lifetimes due to the reduced overlap of the electron-hole wave functions. We demonstrate electric field control of indirect excitons in \HSd hetero-bilayers embedded in a field effect structure with few-layer hexagonal boron nitrite as insulator and few-layer graphene as gate-electrodes. The different strength of the excitonic dipoles and a distinct temperature dependence identify the indirect excitons to stem from optical interband transitions with electrons and holes located in different valleys of the hetero-bilayer featuring highly hybridized electronic states. For the energetically lowest emission lines, we observe a field-dependent level anticrossing at low temperatures. We discuss this behavior in terms of coupling of electronic states from the two semiconducting monolayers resulting in spatially delocalized excitons of the hetero-bilayer behaving like an artificial van der Waals solid. Our results demonstrate the design of novel nano-quantum materials prepared from artificial van der Waals solids with the possibility to in-situ control their physical properties via external stimuli such as electric fields.
\end{abstract}

\maketitle

Van der Waals (vdW) hetero-bilayers (HBs) prepared from optically active transition metal dichalcogenide (TMDC) monolayers, such as \molyd or \woly, combine the excellent properties of the individual layers with the potential for novel functionalities provided by the full control of vdW architecture \cite{Geim.2013, Rivera.2018}. 
In particular, the additional rotational degree of freedom allows to manipulate the electronic coupling between the layers and facilitates the tailoring of the Moiré superlattice potential \cite{Seyler.2018}. By stacking two different TMDC monolayers, the  staggered electronic bands create an atomically sharp p-n heterojunction, effectively separating photo-generated electron-hole pairs. This charge transfer \cite{Zhu.2015b, Chen.2016, Rivera.2018} allows for the formation of so-called interlayer excitons (IXs) with electrons and holes residing in different layers \cite{Rivera.2015, Rivera.2016}. The reduced overlap of the electron-hole wavefunctions entails greatly prolonged exciton lifetimes of more than \SI{100}{\nano\second} \cite{Miller.2017, Rivera.2015}, exceeding the lifetime of intralayer excitons by several orders of magnitude \cite{Robert.2016, Palummo.2015}. The long lifetimes facilitate the formation of thermalized dense exciton ensembles \cite{Nagler.2017}, which, together with IX diffusion over several micrometers \cite{Unuchek.2018, Kulig.2018} sets up the possibility to operate functional excitonic devices by means of electrical manipulation \cite{Unuchek.2018, Jauregui.2018}.
This synergy of fascinating physical properties makes the TMDC HBs not only a promising platform for application in the area of solid state lighting \cite{Withers.2015}, energy conversion \cite{Parzinger.2017}, as well as opto- and valleytronics \cite{Jones.2013, Schaibley.2016}, but also in exciton based information technologies \cite{Unuchek.2018, Ciarrocchi.2018}, and in the design of novel designer quantum nanomaterials. Moreover, vdW structures exhibit potential to access collective excitonic phenomena like high-temperature superfluidity and Bose-Einstein condensation at relaxed experimental conditions \cite{Fogler.2014} as well as excitons in Moiré-superlattices \cite{Seyler.2018}.\par
The coexistence of momentum direct and momentum indirect IXs has been reported to depend crucially on the material combination and their rotational alignment \cite{Miller.2017, Kunstmann.2018, Nayak.2017}. As recent DFT calculations show, different hybridization degrees of the electronic states of the constituent layers can cause the charge carriers to be delocalized across the HB at specific points in the Brillouin zone (BZ), while the carriers at other points in the BZ remain almost fully localized in one of the layers \cite{Gao.2017, Okada.2018, Deilmann.2018, Torun.2018}. Along this line, these structure have to be thought of in terms of `novel artificial vdW solids', more than just of vdW HBs. The hybridization of bands in vdW HBs, a so far almost unexplored field, has not only consequences for excitonic properties but also for transport properties in TMDC-HBs, since in-plane and out-of-plane charge transport might be dominated by completely different electronic states in $k$-space.\par
We experimentally explore the fascinating multi-valley physics in TMDC-HBs and the formation of different types of IXs by investigating their dependence on temperature and applied electric fields and interpret the results according to DFT calculations of the electronic band structure of the \HSd HBs. We achieve electrical control of these distinct IXs in a \HSd HB and find a field-dependent level anti-crossing of the energetically lowest IX emission lines at low temperatures. According to a theoretical suggestion by Gao \textit{et al.} \cite{Gao.2017}, residual coupling of electronic states result in such an anti-crossing behavior of IXs accompanied by a gradual change of the exciton nature from primarily interlayer to primarily intralayer. Our results pave the way for designing novel nano-quantum materials or artificial `vdW solids', with the possibility to in-situ control their physical properties via external stimuli such as electric fields.\par 
\begin{figure}
\includegraphics[scale=1]{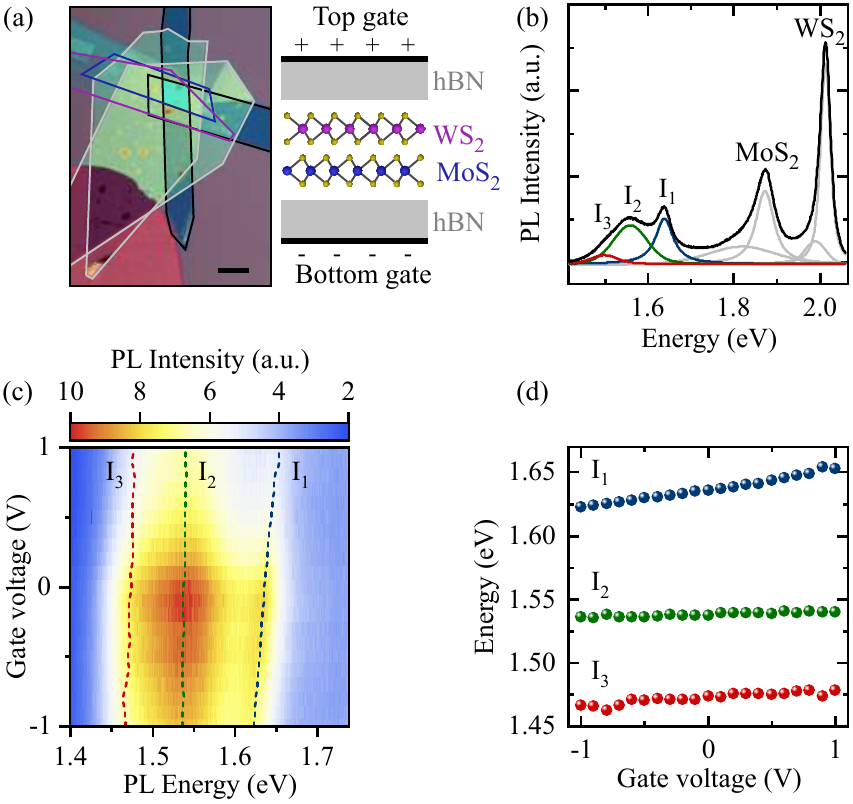}
\caption{(a) Optical microscope image and schematic depiction of the device. The turquoise area indicates the HB area. Scale bar is \SI{10}{\micro\meter}. (b) Room temperature PL spectrum from the HB at zero field, showing emission of \molyd and \wolyd intralayer excitons and IXs $I_1$, $I_2$ and $I_3$. (c) 2D plot of IX PL emission as a function of applied gate voltage, measured at room temperature. Dashed lines are guides to eye indicating the voltage-dependent emission energies of IXs $I_1$, $I_2$ and $I_3$. Positive gate voltages correspond to a positively charged top-gate. (d) Voltage-dependent emission energies of IXs at room temperature.
}
\label{fig: 1}
\end{figure}
The devices prepared for this study consist of \molyd and \wolyd monolayers encapsulated with few-layer hBN and sandwiched between graphite bottom-gate and top-gate electrodes. The HBs are prepared by micromechanical exfoliation from bulk crystals and an all-dry viscoelastic stamping method \cite{CastellanosGomez.2014}. The top monolayer of \wolyd is stacked onto the bottom monolayer of \molyd such that the crystal axes are rotationally aligned to $\ang{0}$ or $\ang{60}$ with a precision of about $\pm\ang{2}$ (see supporting information (SI) for details).  An optical micrograph and a schematic of the device are shown in Figure \ref{fig: 1}a. The field effect area of the HB, where the \molyd and \wolyd monolayers are encapsulated between hBN and graphite layers, is displayed in the center of the image and highlighted in turquoise. A typical photoluminescence (PL) spectrum taken at room temperature using an excitation laser with an excitation energy of E$_\text{laser}=\SI{2.54}{\electronvolt}$ and power of $\text{P}_\text{laser} = \SI{100}{\micro\watt}$, corresponding to an intensity of about $\SI{10}{\kilo\watt\per\centi\meter^2}$ (spot-size ~$\SI{1}{\micro\meter}$), is depicted in Figure \ref{fig: 1}b. The emission bands at higher energies belong to the direct excitons of \wolyd and \moly. The well resolved triplet-structure occurring between \SI{1.4}{\electronvolt} and \SI{1.7}{\electronvolt} that is absent in the monolayer spectra, is interpreted to stem from IXs in agreement with a recent report by Okada \textit{et al.} \cite{Okada.2018}. For a quantitative analysis, we describe the spectrum by a sum of seven Voigt-profiles (cf. SI for further details). The four higher-energy profiles describe the neutral and charged direct excitons in each layer \cite{Chen.2016}, and the three profiles at lower energies labeled as $I_{1}$, $I_{2}$ and $I_{3}$ describe the IX emission. The line-width at room temperature described by the full width half maximum (FWHM) is  around \SI{60}{\milli\electronvolt} for $I_1$ and $I_3$, respectively, and around \SI{125}{\milli\electronvolt} for $I_2$. We would like to note that for a very similar, hence, non-hBN encapsulated \HSd HB, the triplet structure cannot be well resolved (cf. SI). The absence of a well resolved triplet structure reveals the importance of hBN encapsulation for narrow emission bands even at room temperature \cite{Wierzbowski.2017, Cadiz.2017}. Figure \ref{fig: 1}c depicts the evolution of the IXs in the presence of an external electric field at room temperature. The PL intensity (color coded) is plotted as a function of photon energy (bottom axis) and applied gate voltage (left axis). The dashed lines are guides to the eye indicating the voltage-dependent emission energies of IXs $I_1$, $I_2$ and $I_3$, obtained from the fits to the PL spectra at each voltage step. The extracted PL emission energies of the different IX contributions are plotted as a function of gate voltage in Figure \ref{fig: 1}d. Within the considered range of applied gate voltage, the IX contributions exhibit linear Stark shifts $\Delta E$ of different magnitudes. We attribute the non-uniform shifts of the three IX signatures to different out-of-plane dipole moments $\mu_0$. In dependence on the electric field $\boldsymbol{F}$ applied normal to the HB, the Stark shift is given by $\Delta E = \mu_0\boldsymbol{F} + \alpha \boldsymbol{F}^2$, where $\mu_0=ed_0$ is the intrinsic dipole moment, $d_0$ denotes the distance between electron and hole and $\alpha$ is the polarizability of the exciton. Because of the vertical separation of electron and hole, we assume the linear term to be the dominating contribution to the Stark shift. This assumption is in agreement with the experimentally observed linear slopes of the PL peak energies at room temperature with varying gate voltage (see Figure \ref{fig: 1}d). The unequal values of $\mu_{0}$ indicate different separations of the electron-hole pairs along the direction of $\boldsymbol{F}$ and perpendicular to the HB. This is consistent with distinctively different orbital compositions of the electronic states in $k$-space contributing to the excitonic transitions of the \HSd HB. The Stark shifts at room temperature for the three IX transitions are \SI[per-mode=symbol]{15.1}{\milli\electronvolt\per\volt} for $I_1$, \SI[per-mode=symbol]{2.5}{\milli\electronvolt\per\volt} for $I_2$, and \SI[per-mode=symbol]{6.3}{\milli\electronvolt\per\volt} for $I_3$. The estimated separation of the electron and hole  along the electric field constitute approximately \SI{1.8}{\angstrom},  \SI{0.3}{\angstrom} and \SI{0.8}{\angstrom} for $I_1$, $I_2$ and $I_3$, respectively. The direct excitons of each monolayer show small Stark shifts due to their very small dipole moment normal to the layer in agreement with recent reports \cite{Roch.2018}. \par 
\begin{figure}
\includegraphics{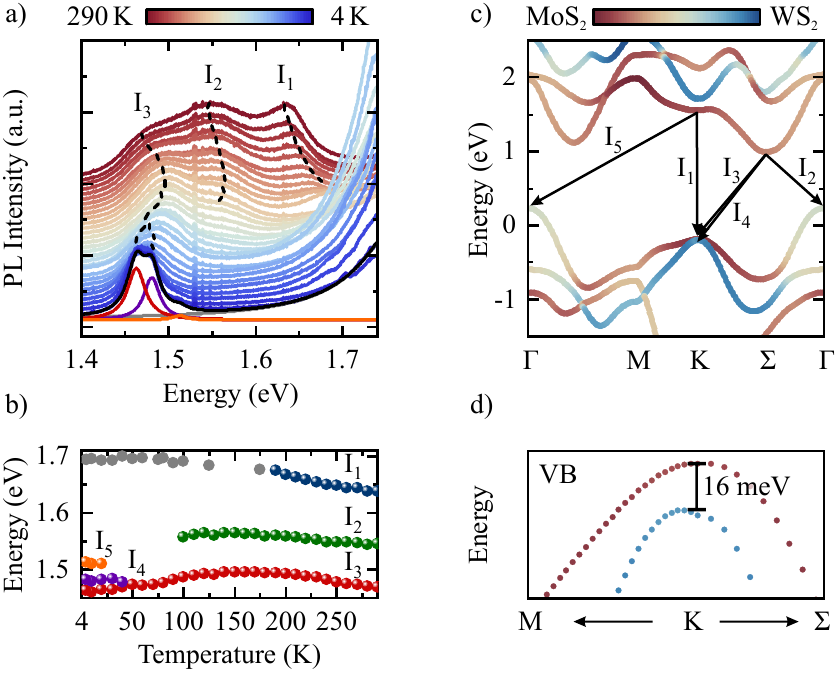}
\caption{(a) IX PL spectra for bath temperatures between \SI{290}{\kelvin} (dark red) and \SI{4}{\kelvin} (dark blue). (b) Emission energies of IXs $I_1$, $I_2$, $I_3$, $I_4$, and $I_5$ as a function of bath temperature. We would like to note that the grey solid dots of the $I_1$ branch are measured on a different HS, where the superimposing low energy tail of the direct exciton emission is less pronounced while the overall signatures from both data sets are very similar. (c) Calculated band structure of a \HSd HB projected onto the atomic orbitals of the individual layers. Red and blue colors correspond to \molyd and \wolyd monolayers and colors in between denote intermediate degree of hybridization. The anticipated excitonic interband transitions are marked by arrows. (d) Magnified view of the VB states at the $K^\text{VB}$-point with a calculated splitting of $\approx\SI{16}{\milli\electronvolt}$.}
\label{fig: 2}
\end{figure}
The evolution of the IX emission with temperatures in the range from \SI{290}{\kelvin} to \SI{4}{\kelvin} is displayed by a waterfall representation in Figure \ref{fig: 2}a. The PL intensity is plotted on a linear scale and the spectra are equidistantly displaced for clarity. The spectra plotted in dark red and dark blue represent temperatures of \SI{290}{\kelvin} and \SI{4}{\kelvin}, respectively. For a quantitative analysis, the spectra are described by an adequate sum of peak-profiles as exemplary shown for the spectrum at the lowest temperature. The temperature dependence of the peak positions is summarized in Figure \ref{fig: 2}b, revealing that the IX contributions evolve differently with temperature. As expected from an increasing single particle band gap with decreasing temperature, the emission energy of $I_1$ increases monotonously with temperature, similar to the direct excitons (not shown). The emission energy of $I_2$ is almost constant and the intensity continuously decreases with decreasing temperature, becoming indistinguishable from the noise level below \SI{100}{\kelvin}. The emission energy of $I_3$ has a very broad maximum around \SI{150}{\kelvin}. The \SI{4}{\kelvin} spectrum in Figure \ref{fig: 2}a clearly demonstrates, that the line splits into a well resolved doublet structure labeled with $I_3$ and $I_4$ at temperatures below \SI{40}{\kelvin}. The intensity of $I_4$ increases by a factor of two from \SI{40}{\kelvin} to \SI{4}{\kelvin}. Notably, below \SI{30}{\kelvin} another weak emission line $I_5$ appears energetically just above $I_4$.
The values for the stark shifts $\Delta E$ at \SI{4}{\kelvin} are very similar to the values observed at room temperatures and are \SI[per-mode=symbol]{17.0}{\milli\electronvolt\per\volt} for $I_1$ and $\approx\SI[per-mode=symbol]{7} {\milli\electronvolt\per\volt}$ for $I_3$, $I_4$ and $I_5$.
\begin{table}[b]
	\resizebox{\columnwidth}{!}{%
		\begin{tabular}{llll}
			\hline\hline
			$k$-space & Majority of Orbitals & Minority of Orbitals & Hybridization $\text{\HS}$ \\ \hline
			$K^\text{CB}$ & $M$-$d_{z^2}$ & $X$-$p_x,p_y$ & $\percent{90}$ $\text{\moly}$, $\percent{10}$ $\text{\woly}$ \\
			$K^\text{VB1}$ & $M$-$d_{x^2-y^2},d_{xy}$ & $X$-$p_x,p_y$ & $\percent{85}$ $\text{\moly}$, $\percent{15}$ $\text{\woly}$ \\
			$K^\text{VB2}$ & $M$-$d_{x^2-y^2},d_{xy}$ & $X$-$p_x,p_y$ & $\percent{10}$ $\text{\moly}$, $\percent{90}$ $\text{\woly}$ \\
			$\mathit{\Sigma}^\text{CB}$ & $M$-$d_{x^2-y^2},d_{xy}$ & $M$-$d_{z^2}$, $X$-$p_x,p_y,p_z$ & $\percent{75}$ $\text{\moly}$, $\percent{25}$ $\text{\woly}$ \\
			$\mathit{\Gamma}^\text{VB}$ & $M$-$d_{z^2}$ & $X$-$p_z$ & $\percent{40}$ $\text{\moly}$, $\percent{60}$ $\text{\woly}$ \\ \hline\hline
		\end{tabular}%
	}
	\caption{Orbital compositions of Bloch states at critical $k$-space symmetry points $K^\text{CB}$, $\mathit{\Sigma}^\text{CB}$, $K^\text{VB}$ and $\mathit{\Sigma}^\text{VB}$ in conduction band (CB) and valence band (VB) of $MX_2$ compounds \cite{Feng.2012, Chang.2013, Liu.2015}. Hybridization of electronic states in a \HSd HB is extracted from DFT calculations.}
	\label{tab: 1}
\end{table}
Overall, the peculiar temperature dependence and significantly different Stark shifts strongly suggest that the IXs are formed by electrons and holes residing in different valleys of the BZ. The lowest conduction bands (CB) and topmost valence bands (VB) of the \HSd HB plotted in Figure \ref{fig: 2}c are calculated by density functional theory (DFT) 
and agree reasonably well with recent literature \cite{Deilmann.2018}. 
The suggested interband transitions of the HB and the assignment to the experimentally observed emission lines $I_1$ - $I_5$ are also depicted by arrows in Figure \ref{fig: 2}c. The single particle bands are projected onto the atomic orbitals of the individual layers that are reflected by the color code.  Red and blue colors correspond to electronic states purely from \molyd and \woly, respectively, whereas intermediate colors indicate hybridization of states from both layers. The hybridization is a consequence of coupling between electronic states on the molecular-orbital level that is particularly strong at the $\mathit{\Sigma}$ valley in the CB and the $\mathit{\Gamma}$ valley in the VB. The hybridization of the electronic states for the \HSd HB at the relevant high symmetry points in $k$-space extracted from the DFT calculus are summarized in Table \ref{tab: 1} and are in excellent agreement with recent literature \cite{Gao.2017,Deilmann.2018,Torun.2018}. Interestingly, the lowest CB states at the $K^\text{CB}$-point are of \percent{90} \molyd character, whereas the two topmost VB states at the $K^\text{VB}$-point are splitted by only $\approx \SI{16}{\milli\electronvolt}$ and are of inverse orbital character with \percent{85} \molyd and \percent{90} \wolyd for the higher and lower state, respectively (see Figure \ref{fig: 2}d). Similar to TMDC homo-bilayers, also the \HSd HBs are indirect semiconductors with the lowest energy transition between the  $\mathit{\Sigma}^\text{CB}$ and the $K^\text{VB}$ valley. An interesting aspect of the high hybridization of the electronic states for some high symmetry points of the \HSd HB is the fact that electrons and holes residing in these valleys are delocalized between the two constituent layers and hence behave more like charge carriers of a new `artificial vdW solid' than of a vdW HB. Taking into account this aspect, we assign the emission line of $I_2$, having the smallest Stark shift (at room temperature) and hence the smallest separation between electron-hole pairs, to be formed by an electron at $\mathit{\Sigma}^\text{CB}_\text{HB}$ and a hole at $\mathit{\Gamma}^\text{VB}_\text{HB}$. The small but finite vertical separation of the charge carriers results from a large degree of orbital hybridization with \percent{75} \molyd character at the $\mathit{\Sigma}^\text{CB}$-point and \percent{60} \wolyd character at the $\mathit{\Gamma}^\text{VB}$ point. Nevertheless, $I_2$ can be interpreted as momentum indirect, but real space more direct transition and, hence, as exciton of the \HSd vdW solid.
With similar arguments, we can assign $I_1$, which shows the largest Stark shift and hence the largest electron-hole separation to be a spatially indirect exciton, such that the electron is located in the almost fully layer polarized $K^\text{CB}_\text{\moly}$-point ($\approx$ \percent{90} \moly) and the hole is located in the layer polarized $K^\text{VB2}_\text{\woly}$-point ($\approx$ \percent{90} \woly). The $I_1$ line is therefore from a momentum direct, but spatially indirect exciton. The remaining excitons $I_3$, $I_4$ and $I_5$ exhibit all similar Stark shift values that are between those of $I_1$ and $I_2$. For this reason, it is most likely that for these interband transitions one of the charge carriers is delocalized between the two layers and resides consequently in a highly hybridized valley, while the other Coulomb coupled charge carrier is localized in either \molyd or \wolyd and belongs therefore to a non- or only weakly hybridized valley. Taking into account the energies of the IXs with  $E_\text{I5} > E_\text{I4} > E_\text{I3}$, we assign $I_5$ to the transition with the electron at $K^\text{CB}_\text{\moly}$ and the hole at the hybridized $\mathit{\Gamma}^\text{VB}_\text{HB}$-point. The transitions $I_3$ and $I_4$ are assigned to transitions with the electron residing at the hybridized $\mathit{\Sigma}^\text{CB}_\text{HB}$ valley for both transitions and the hole at the $K^\text{VB1}_\text{\moly}$-point and the $K^\text{VB2}_\text{\woly}$-point, respectively. The $K^\text{VB1}_\text{\moly}$ valley is energetically located above the $K^\text{VB2}_\text{\woly}$ valley. The experimentally found energy separation of the two emission lines is about \SI{20}{\milli\electronvolt} in excellent agreement with the theoretical value of about \SI{16}{\milli\electronvolt} obtained from the DFT calculation (Figure \ref{fig: 2}d). The three excitons $I_3$, $I_4$ and $I_5$ are momentum indirect IXs between the HB and one of the constituent TMDC monolayers. In brief, the five red-shifted emission lines are assigned to a momentum direct interlayer transition $I_1$ from $K^\text{CB}_\text{\moly}$ to $K^\text{VB2}_\text{\woly}$, a  momentum indirect intralayer transition $I_2$ from $\mathit{\Sigma}^\text{CV}_\text{HB}$ to $\mathit{\Gamma}^\text{VB}_\text{HB}$ and three momentum indirect interlayer transitions, namely $I_3$ from $\mathit{\Sigma}^\text{CB}_\text{HB}$ to  $K^\text{VB1}_\text{\moly}$, $I_4$ from $\mathit{\Sigma}^\text{CB}_\text{HB}$ to $K^\text{VB2}_\text{\woly}$ and $I_5$ from $K^\text{CB}_\text{\moly}$ to $\mathit{\Gamma}^\text{VB}_\text{HB}$.\par
The increase of the emission energy of $I_3$ with decreasing temperature between \SI{290}{\kelvin} and \SI{150}{\kelvin} is consistent with an increase of the single particle band gap of both constituent layers \cite{Varshni.1967}. The subsequent lowering of the emission energy below \SI{150}{\kelvin} might be caused by reduction of the amount of thermally excited kinematic excitons that can couple to the light, since we expect a slight lattice mismatch to cause a misalignment in $k$-space \cite{Miller.2017, Yu.2015}. The phonon-activated transition $I_2$ is rather broad and disappears below about \SI{100}{\kelvin}, which is expected to be caused by the relevant phonon mode freezing out and a reduced electron-phonon coupling.
The increase of the intensity of $I_3$, $I_4$ and $I_5$ with decreasing temperature goes along with the decrease of the intensity of $I_1$ and is consistent with a reduction of the momentum direct recombination rate at the $K$-points due to a reduced thermal broadening of the quasi Fermi-energies \cite{Miller.2017, Yu.2015}. Similarly, the fact that the separation into three lines can only be well resolved at temperatures below \SI{40}{\kelvin} can be explained by a reduced thermal broadening of the levels and reduced scattering.
\begin{figure}
\includegraphics{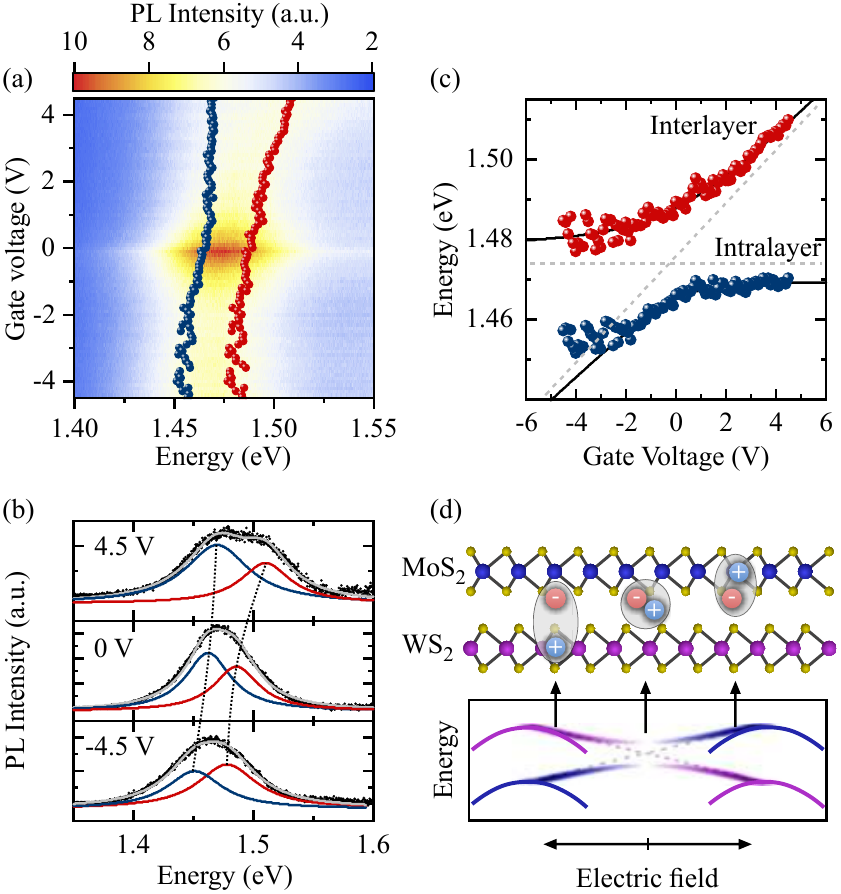}
\caption{(a) 2D plot of exciton PL emission as a function of applied gate voltage measured at a bath temperature of \SI{10}{\kelvin}. Blue and red dots are guides to the eye indicating the gate voltage-dependent evolution of excitonic emission. (b) PL spectra at applied gate voltages of \SI{-4.5}{\volt}, \SI{0}{\volt} and \SI{4.5}{\volt}. (c) Emission energy of upper and lower exciton branch as a function of applied gate voltage. The gray dashed lines indicate the interlayer and intralayer character of the exciton emission. (d) Schematic depiction of the anti-crossing behavior of the valence band states at the $K^\text{VB}$-point with varying electric field, leading to interlayer-like and intralayer-like excitons.}
\label{fig: 3}
\end{figure}
It is peculiar that the topmost VB states $K^\text{VB1}_\text{\woly}$ are \percent{85} of \molyd orbitals, whereas the second topmost VB states $K^\text{VB2}_\text{\woly}$ are \percent{90} of \wolyd orbitals. The hole contributing to the interband transition $I_3$ is therefore almost fully localized in \moly, whereas the hole contributing to the transition $I_4$ is almost fully localized in \woly. For both transitions, the electron is more delocalized between both layers with an orbital projection of \percent{75} on \molyd at the lowest CB states $\mathit{\Sigma}^\text{CB}_\text{HB}$. Stark effect measurements at $\text{T} = \SI{10}{\kelvin}$  in a large electric field range from \SI[per-mode=symbol]{-2.2}{\mega\volt\per\centi\meter} to \SI[per-mode=symbol]{+2.2}{\mega\volt\per\centi\meter}, realized by tuning the the gate voltage from \SI{-4.5}{\volt} to \SI{+4.5}{\volt}, reveal a clear anti-crossing behaviour of the emission lines stemming from $I_3$ and $I_4$, as demonstrated in Figure \ref{fig: 3}. The gate voltage-dependent evolution of the $I_3$ and $I_4$ PL signal is plotted in Figure \ref{fig: 3}a, where the intensity is color coded and the dots highlight the peak-position from fits to the spectra using two peak profiles. Exemplary spectra for gate voltages of \SI{-4.5}{\volt}, \SI{0}{\volt} and \SI{4.5}{\volt} are displayed in Figure \ref{fig: 3}b. It is clearly visible that each spectrum needs to be described by two individual peak profiles. The peak positions of the two emission bands $I_3$ and $I_4$ are plotted as a function of gate voltage in Figure \ref{fig: 3}c, demonstrating the gate-dependent anti-crossing behaviour of $I_3$ and $I_4$. The energetically lower emission branch $I_3$ exhibits a larger Stark shift for negative gate voltages, that gradually decreases towards positive gate voltages and almost saturates for large positive gate voltages. In contrast, the Stark shift of the energetically higher branch $I_4$ gradually increases towards positive gate voltages. We interpret the anti-crossing behaviour following the arguments of Gao \textit{et al.} \cite{Gao.2017} in terms of finite interlayer coupling for the topmost and the second topmost valence band states $K^\text{VB1}_\text{\moly}$ and $K^\text{VB2}_\text{\woly}$. The projection of the electronic wave function on either layer at $K^\text{VB1}_\text{\moly}$ is not purely \woly, but contains \percent{15} of \woly, and $K^\text{VB2}_\text{\woly}$ contains \percent{10} of \moly. Under the action of an electric field $\boldsymbol{F}$, the relative band alignment of the two layers responds linearly as 
\begin{equation*}
    E_\text{\moly}(F)-E_\text{\woly}(F) = E_\text{\moly}(0) - E_\text{\woly}(0) -eFd/\varepsilon,
\end{equation*}
where $E_\text{\moly}$ and $E_\text{\woly}$ are the band energies of the single layers, $Fd$ describes the voltage drop across the HB and $\varepsilon$ is the effective dielectric function of the materials, describing its screening response \cite{Gao.2017}. Without finite interlayer coupling, the valence band states of the two bands at $K$ would pass through each other. The finite coupling of the VB states at $K$, however, causes level repulsion which manifests in an avoided crossing \cite{Gao.2017} as sketched in Figure \ref{fig: 3}d. We would like to note that the electronic states at $\mathit{\Sigma}^\text{CB}_\text{HB}$ that are involved in these interband transitions are asymmetrically delocalized between the \molyd and \wolyd monolayers, with a higher contribution of \molyd (\percent{75}) than of \wolyd (\percent{25}). 
The solid line plotted in Figure \ref{fig: 3}c describes the coupling strength $\gamma$ using a simplified model of a coupled two-level system
\begin{equation*}
    E_\pm =\frac{1}{2}(\Delta(0)-eFd/\varepsilon)\pm\sqrt{(\Delta(0)-eFd/\varepsilon)^2+4\gamma^2},
\end{equation*}
where $E_\pm$ denotes the upper (+) and lower (-) energy branch and $\Delta(0)=E_\text{\moly}(0)-E_\text{\woly}(0)$ \cite{Gao.2017}. The model fit to the data yields $\gamma_+ = \SI{11}{\milli\electronvolt}$ and $\gamma_-=\SI{5}{\milli\electronvolt}$ for the upper branch $I_4$ and the lower branch $I_3$, respectively. We assume the asymmetry to result from the fact that the electron states at $\mathit{\Sigma}^\text{CB}_\text{HB}$ are \percent{75} in \molyd and \percent{25} in \woly. As a direct consequence of the avoided crossing of the VB states at the $K$-point, the excitons of two emission lines $I_{3}$ ($I_{4}$) gradually change their nature from interlayer  with the delocalized electron at $\mathit{\Sigma}^\text{CB}_\text{HB}$ and the hole at $K^\text{VB1}_\text{\moly}$ ($K^\text{VB2}_\text{\woly}$), to intralayer in the anti-crossing regime with delocalized electron at $\mathit{\Sigma}^\text{CB}_\text{HB}$ and delocalized hole at $K^\text{VB}_\text{HB}$. Further increasing the positive gate voltage gradually turns the nature of the exciton back to an IX with the delocalized electron unchanged at $\mathit{\Sigma}^\text{CB}_\text{HB}$ and the localized hole at $K^\text{VB2}_\text{\woly}$ ($K^\text{VB1}_\text{\moly}$). Remarkably, the excitons of $I_{3}$ and $I_4$ do not only change their nature from interlayer to intralayer in the anti-crossing regime, but the hole also changes its hosting layers. For instance at the emission line $I_3$, the hole is localized in \wolyd for large negative gate voltages and localized in \molyd for large positive gate voltages and vice versa for $I_4$, as sketched in Figure \ref{fig: 3}d. The change of the nature of $I_3$ and $I_4$ manifests in the gate voltage-dependent change of the Stark shift indicating altered out-of-plane dipole moments that are minimal for the hole \percent{85} localized in \moly, and largest for the hole \percent{90} localized in \woly, while the electron remains at a layer projection of \percent{75} \moly. 
\begin{figure}
\includegraphics{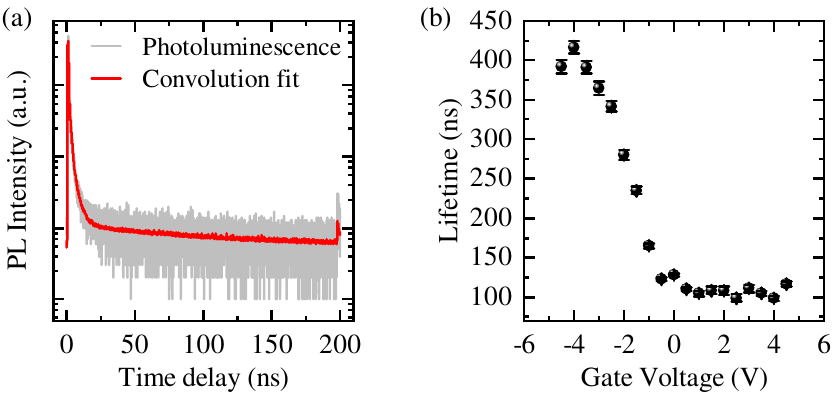}
\caption{(a) PL signal of IXs as a function of time delay, measured at a bath temperature of \SI{10}{\kelvin} and \SI{-4}{\volt} gate voltage. The gray line is experimental data and the red line is a fit by a triexponential function convoluted by the instrument response. (b) Extracted PL lifetime of the longest decay component as a function of applied gate voltage.}
\label{fig: 4}
\end{figure}
We expect therefore the lifetime of the emission to depend strongly on the gate voltage since a reduced overlap of the electron and the hole wave-function increases the exciton lifetime. Following this argument, we expect the longest lifetime of $I_3$ for negative gate voltages, gradually decreasing with increasing gate voltage and saturating at large positive gate voltages with the hole being transferred from \wolyd to \moly, going from negative to positive gate voltages. An exemplary time-resolved PL trace of the combined $I_3$ and $I_4$ emission line (emission energies below \SI{1.65}{\electronvolt}) at $\text{T}=\SI{10}{\kelvin}$ and a gate voltage of \SI{-4}{\volt} is plotted in Figure \ref{fig: 4}a. From exponential fits to the data, we extract rather long decay times in the order of \SI{100}{\nano\second} to \SI{400}{\nano\second}. We attribute this long-lived decay channel to be dominated by the $I_3$ emission that is brighter compared to the $I_4$ emission and is expected to have the longest lifetime of all IXs as discussed above. Indeed, the extracted lifetime is about \SI{400}{\nano\second} for negative gate voltages, decreases to about \SI{100}{\nano\second} for zero gate voltage and saturates for larger gate voltages as shown in Figure \ref{fig: 4}b. We attribute the saturation in lifetime to the fact that the overlap of electron and hole wave-functions is very similar in the hybridized regime (small gate voltages) and for the hole localized in \molyd (positive gate voltages), whereas the overlap is reduced for the hole localized in \wolyd due to the fact that the electron is delocalized with \percent{25} \wolyd and \percent{75} \molyd (negative gate voltages).  The lifetime measurements support our interpretation of the anti-crossing behaviour of the emission lines $I_3$ and $I_4$. As sketched in Figure \ref{fig: 3}c, by changing the gate voltage and hence the electric field normal to the layers, the nature of the excitons changes gradually from interlayer to intralayer and back to interlayer with the hole being transferred from one layer to the other one by changing the direction of the electric field $\boldsymbol{F}$, realized by changing the polarity of the bias voltage.\par
To conclude, we demonstrate widely tunable IX emission by applying an electric field perpendicular to the HB plane. Field-dependent PL measurements at room temperature reveal distinctive energy shifts of IXs with varying electric field, which we interpret to result from different orbital hybridization within the excitonic $k$-space transitions. The control of the exciton energy under application of an electric field via the quantum confined stark effect (QCSE) \cite{Klein.2016, Mak.2018, Ciarrocchi.2018, Wang.2018, Roch.2018} can be applied to trap excitons in vdW HBs, to create and manipulate dense exciton ensembles \cite{Gartner.2007, Dietl.2017} mandatory to study many body correlation physics of these composite bosons, to create single excitons or exciton molecules \cite{Schinner.2013}, and to implement functional exciton devices such as exciton transistors \cite{Kuznetsova.2010, Unuchek.2018}. Furthermore, we discuss anti-crossing behaviour in field-dependent time-integrated and time-resolved PL measurement of the lowest IX emission lines. The experimental findings suggest a field-dependent change of the exciton nature from interlayer to intralayer and back to interlayer, accompanied with the transfer of the participating hole from one constituent layer to the other one. Interestingly, for small positive or negative gate voltages, both, electron and hole states are delocalized between the \molyd and \wolyd layers due to interaction driven hybridization of the single particle bands. This observation manifests the fact that vdW HBs need to be treated as artificial solids with the band-structure of the HB and not just as heterostructures with the combination of the isolated single particle bands of the constituent layers. Another visionary aspect hereby is, that in-plane and cross-plane transport in these structures is expected to occur in different valleys, where it is unlikely that in-plane and out-of-plane transport channels can cross-talk. In turn, they can be seen as topological protected paths for charge transport paving the ground for new technological concepts in the realization of innovative architecture in two-dimensional electronic circuitries.\\
\par

\paragraph{Author contributions}
J.K and F.S. contributed equally to this work. J.K, F.S. and B.M. performed the measurements. J.K and F.S. prepared the samples. K.W. and T.T provided high-quality hBN. U.W. and A. H. conceived the experiment. M.L. performed the DFT caclulation. J.K., F.S., A.H. and U.W. analyzed the data. J.K. and U.W. prepared the Figures and wrote the manuscript. All authors discussed the results and commented on the manuscript.\\
\par
\paragraph{Acknowledgements}
We gratefully acknowledge financial support by the Deutsche Forschungsgemeinschaft (DFG) via excellence cluster `Nanosystems Initiative Munich' (NIM) and DFG projects WU 637/4- 1 and HO 3324/9-1. M.L. was supported by the Deutsche Forschungsgemeinschaft (DFG) within RTG 2247 and through a grant for CPU time at the HLRN (Hannover/Berlin). K.W. and T.T. acknowledge support from the Elemental Strategy Initiative conducted by the MEXT, Japan and and the CREST (JPMJCR15F3), JST.

\section{References}

\clearpage

\end{document}